\documentclass[aps,pra,twocolumn,superscriptaddress,showpacs,showkeys,amsmath,amssymb]{revtex4}
\usepackage{amsfonts}
\usepackage{amssymb,amsmath}
\usepackage{mathrsfs}
\usepackage{latexsym}
\usepackage{amsmath}
\usepackage[cp1251]{inputenc}
\usepackage{graphicx}
\usepackage{dcolumn}
\usepackage{bm}
\usepackage{color}

\RequirePackage{ifthen}
\RequirePackage[pdfstartview=FitH]{hyperref}
\begin{document}
	
	\title{Two- and three-body bound states in one-dimensional Fermi bipolaron\\ with three-body interaction}
	\author{O.~Hryhorchak\footnote{e-mail: hrorest@gmail.com}} \affiliation{Professor Ivan Vakarchuk Department for Theoretical Physics, Ivan Franko National University of Lviv, 12 Drahomanov Street, Lviv, Ukraine}
	\author{V.~Pastukhov\footnote{e-mail: volodyapastukhov@gmail.com}}
	\affiliation{Professor Ivan Vakarchuk Department for Theoretical Physics, Ivan Franko National University of Lviv, 12 Drahomanov Street, Lviv, Ukraine}

	\date{\today}

	
	
	\begin{abstract}
We discuss the ground-state properties of two bosonic (or spin-1/2 fermionic in a singlet spin state) impurities immersed in a one-dimensional ideal Fermi gas with only the three-body contact interaction accounted for. Despite its simplicity, the considered model is found to demonstrate a variety of medium-induced few-body bound states. Particularly, using variational calculations with trial wave functions that correctly take into account one particle-hole excitation, we predict the emergence of a dimer state and several trimer states over a wide range of the three-body coupling parameter.
	\end{abstract}
	
	\maketitle
\section{Introduction}
The progress in the experimental realization \cite{Schirotzek_2009} of fermions with a small number of impurities in ultra-cold atomic gases stimulated extensive theoretical studies of Fermi polarons \cite{Massignan_2014} in the last two decades. Although this is a foundational model of condensed matter physics with numerous practical applications that have been well studied previously \cite{Mahan_2000}, the possibilities that ultra-cold atomic setups provide for manipulating the two-body interaction through Feshbach resonances \cite{Chin_2010} are tremendous. In particular, one can observe \cite{Kohstall_2012,Koschorreck_2012,Scazza_2017} the repulsive and attractive branches of Fermi polarons, explore peculiarities of the molecule-polaron transition \cite{Punk_2009,Bruun_2010,Parish_2011,Schmidt_2011,Cui_2020,Ness_2020} and even higher few-body complexes \cite{Parish_2013,Liu_2024}. Although its Hamiltonian includes only a contact pairwise potential between impurity and non-interacting host fermions, the model can be exactly solved \cite{McGuire_1965,McGuire_1966,Orso_2026} only in one dimension by means of the Bethe ansatz for equal masses of impurity and host fermions. Except for the static (infinite-mass) impurity limit (see \cite{Schmidt_2018} for review), numerical--Monte Carlo \cite{Pilati_2008,Prokofev_2008,Houcke_2020} methods or some approximate approaches (mostly variational) \cite{Chevy_2006,Combescot_2007,Combescot_2008} should be adopted in all higher dimensions. Despite the simplicity of the trial wavefunction--only one particle-hole excitation is typically included--the parameters of the calculated low-lying spectrum of the Fermi polaron are well reproduced even at unitarity. The polaron problem is known to suffer from Anderson's orthogonality catastrophe \cite{Anderson_1967}, in which the quasiparticle residue decreases as the mass of the impurity atom increases. This peculiarity is correctly captured by methods \cite{Kain_2017,Chen_2025} based on the celebrated Lee-Low-Pines transformation \cite{Lee_1953} or requires a massive resummation \cite{Pimenov_2018} of the Feynman diagrammatic series.

In typical experimental setups, one usually deals with a small number of external particles rather than a single impurity. In this situation, the medium-induced effective interaction \cite{Mistakidis_2019,Enss_2020,Tajima_2021} between polarons, which is generally complex-valued \cite{Akamatsu_2024} due to decoherence effects and strongly depends \cite{Baroni_2024,Paredes_2024,Ando_2025,Levinsen_2025} on impurity statistics, can provide the formation of Fermi bipolarons. In the spin-$1/2$ realization of the bipolaronic system, this induced interaction is weakened by the Pauli principle, provided that the $p$-wave channel \cite{Nishida_2009,Giraud_2012} is the first attractive one, which is necessary for bipolaron formation. For bosonic \cite{Bellotti_2016,Huber_2019,Mukherjee_2020} or indistinguishable atoms immersed in a Fermi sea, the physics is richer, even supporting the formation \cite{Guo_2024} of clusters. The induced interaction stabilizes the formation of composite fermions in three-component Fermi systems \cite{Nishida_2015,Kirk_2017} and in Bose-Fermi mixtures \cite{Hryhorchak_2023,Pastukhov_2024,Pisani_2025} at a strong enough boson-fermion attraction. For low densities of impurity atoms of both statistics, the universal formula for the interaction energy based on the Fermi Liquid theory can be obtained (see recent review Ref.~\cite{Scazza_2022} for details). Effects of induced polaron-polaron interaction are found \cite{Hu_2018,Tajima_2018} to be sensitive to increasing temperature.

Here, we address the problem of two non-identical impurities immersed in a one-dimensional non-interacting fermion system with all two-body potentials fine-tuned to resonance, and only the three-body interaction preserved. Such a composition with only two distinct impurities in the Fermi sea effectively reduces the problem to the Fermi-polaronic one, which can be treated with well-established approximate methods, particularly the celebrated Chevy ansatz \cite{Chevy_2006}. In the last decade, low-dimensional systems of particles interacting via a contact three-body potential attracted much attention in the literature, both in few- \cite{Guijarro_2018,Nishida_2018,Valiente_2019,McKenney_2019,Sekino_2021,Hryhorchak_2022,Polkanov_2024,Polkanov_2026}, and many-body \cite{Sekino_2018,Pastukhov_2019,Maki_2019,Czejdo_2020,McKenney_2020,Akagami_2021,Morera_2022,Tajima_2022,Tajima_2025,Polkanov_2026_2} contexts. This is the most relevant (in the renormalization-group sense) perturbation in two spatial dimensions or fewer. In a one-dimensional setup, it not only breaks integrability but also exhibits an intrinsic scale anomaly \cite{Drut_2018}. The latter affects the many-body properties of the system \cite{Valiente_Pastukhov,Tajima_2024}, enabling its experimental detection.

\section{Formulation}
\subsection{Model}
The model under consideration comprises macroscopically populated non-interacting fermions of mass $m$ in one dimension (1D) and two bosonic (or fermionic in an antisymmetric spin state) impurities of different masses $m_I$. The system is placed in a large 1D volume $L$ with periodic boundary conditions. Only the three-body zero-range interaction, which simultaneously involves the host fermion and two impurities, is accounted for. The mixed-representation (the first quantized for exterior particles and the second quantized for host fermions) Hamiltonian reads
\begin{eqnarray}\label{H}
&&H=\sum_{j=1,2}\frac{p_j^2}{2m_I}+
\sum_{p}\varepsilon_p\psi^{\dagger}_p\psi_p\nonumber\\
&&+\int dx g_{3,\Lambda}\delta_{\Lambda}(y_1-x)\delta_{\Lambda}(y_2-x)n(x),
\end{eqnarray}
where $x,y_j\in [0,L)$ and $\varepsilon_p=\frac{p^2}{2m}$. The creation $\psi^{\dagger}_p$ and annihilation $\psi_p$ operators obey the standard fermionic commutation relations and $n(x)=\psi^{\dagger}(x)\psi(x)$ denotes the local density of fermions. The second term in $H$ describes the three-body short-range interaction with bare coupling $g_{3,\Lambda}$ that depends on the ultraviolet (UV) cutoff $\Lambda$. The delta-function, $\delta_{\Lambda}(x)=\frac{1}{L}\sum_{|p|<\Lambda}e^{ipx}$, in Eq.~(\ref{H}) is assumed to be smoothed over the scale $\Lambda^{-1}$. All observables below are conventionally \cite{Drut_2018,Pastukhov_2019,Valiente_Pastukhov} related to the three-body vacuum bound-state energy, $\epsilon_3$, via
\begin{eqnarray}\label{g_3}
-g^{-1}_{3,\Lambda}=\frac{1}{L^2}\sum_{|p|, |p'|<\Lambda}\frac{1}{\varepsilon_I(p)+\varepsilon_I(p')+\varepsilon_{p+p'}+|\epsilon_3|},
\end{eqnarray}
where $\varepsilon_I(p)=\frac{p^2}{2m_I}$ is the dispersion relation of a free impurity.

\subsection{Effective Hamiltonian}
A local character of the three-body interaction considered here suggests the partial reduction of the Hilbert space of the initial problem. Particularly, by passing to the center-of-mass $Y=\frac{y_1+y_2}{2}$ and relative motion $y=y_1-y_2$ coordinates, we can rewrite the Hamiltonian as follows
\begin{eqnarray}\label{H_2}
H=\frac{p^2}{m_I}+H_0+\delta_{\Lambda}(y)\int dx g_{3,\Lambda}\delta_{\Lambda}(Y-x)n(x),
\end{eqnarray}
where $H_0=\frac{P^2}{4m_I}+\sum_{p}\varepsilon_p\psi^{\dagger}_p\psi_p$ stands for the kinetic energy of host fermions and the center-of-mass motion of two impurities. Now, it is clear that the degree of freedom describing the relative motion of exterior particles can be traced out. Indeed, the original Schr\"odinger equation $H|\Psi\rangle_{\textrm{full}}=E|\Psi\rangle_{\textrm{full}}$ for the full eigenstate can be projected on the plane-wave basis ($|p\rangle$) in the relative-motion subspace
\begin{eqnarray}
&&\left\{2\varepsilon_I(p)+H_0-E\right\}\langle p|\Psi\rangle_{\textrm{full}}\nonumber\\
&&+g_{3,\Lambda}n(Y)\frac{1}{L}\sum_{p'}\langle p'|\Psi\rangle_{\textrm{full}}=0.
\end{eqnarray}
The formal solution to the above equation, written down in operator form, reads
\begin{eqnarray}\label{Psi}
\langle p|\Psi\rangle_{\textrm{full}}=\frac{\Pi^{-1}(E-H_0)}{2\varepsilon_I(p)+H_0-E}|\Psi\rangle,
\end{eqnarray}
where $|\Psi\rangle$ lives in the reduced Hilbert space of macroscopically-populated fermions plus a center-of-mass motion of impurities, and $\Pi(E)=\frac{1}{L}\sum_{p}\frac{1}{2\varepsilon_I(p)-E}$. Formally, vector $|\Psi\rangle$ is the eigenstate (with eigenvalue $E$) of the Schr\"odinger equation $H_{\textrm{eff}}|\Psi\rangle=E|\Psi\rangle$ with the effective Hamiltonian
\begin{eqnarray}\label{H_eff}
H_{\textrm{eff}}-E\propto \Pi^{-1}(E-H_0)+g_{3,\Lambda}n(Y)
\end{eqnarray}
that describes the center-of-mass motion of two impurities interacting with the host fermions through the effective two-body potential $\Phi_{\textrm{eff}}\propto g_{3, \Lambda}\delta_{\Lambda}(Y-x)$. The obtained $H_{\textrm{eff}}$ is a starting point for further variational analysis of the dimer-trimer transition in the system. 

\section{Variational treatment}
In the following, we adopt a simple ansatz that includes up to one particle-hole excitation of host fermions, both for dimer and trimer states. This simple variational consideration is known \cite{Chevy_2006} to be very efficient for the standard Fermi-polaron problem, where it accurately captures properties of the system even in the unitary limit.

\subsection{Dimer state}
The dimer state of two impurities interacting through the three-body potential with a macroscopic number of 1D fermions was previously analyzed in Ref.~\cite{Hryhorchak_2026} by means of a diagrammatic approach. For convenience, we reproduce these calculations using the effective Hamiltonian (\ref{H_eff}). Additionally, this will establish the connection between our variational approach and $T$-matrix approximation. The appropriate trial wave function that describes the state with a nonzero center-of-mass momentum $P$ of two exterior particles in a Fermi sea of $N$ non-interacting fermions, reads
\begin{eqnarray}\label{dimer}
|D\rangle=de^{iPY}|N\rangle+\frac{1}{L}\sum_{k,q}d_{k;q}e^{i(P-k+q)Y}\psi^{\dagger}_k\psi_q|N\rangle,
\end{eqnarray}
where $|N\rangle$ is the ground state of host fermions without impurities, $|k|>p_F$ and $|q|\le p_F$ ($p_F$ is the Fermi momentum) and coefficients $d$ and $d_{k;q}$ are the subject of $\langle D|H_{\textrm{eff}}|D\rangle$ minimization. An expectation value of the effective Hamiltonian can be easily calculated; therefore, the details are omitted. The forthcoming application of the minimization procedure yields two coupled integral equations (see Appendix A) for the coefficients $d$ and $d_{k;q}$ in the ansatz (\ref{dimer}). The requirement of a non-trivial solution reduces to a single transcendental equation in the limit $\Lambda \to \infty$ (generalized to the $m_I\neq m$ case at zero $P$):
\begin{eqnarray}\label{dimer_Eq}
\Pi^{-1}(\mathcal{E})+\frac{1}{L}\sum_{q}\mathcal{T}_{0;q}(\mathcal{E})=0,
\end{eqnarray}
where $\mathcal{E}=E-\sum_{q}\varepsilon_q$ is the dimer energy, and here (and for later convenience) we define a function
\begin{eqnarray}\label{Tau_3}
\mathcal{T}^{-1}_{k;q}(\mathcal{E})=g^{-1}_{3,\Lambda}+\frac{1}{L}\sum_{k'}\Pi_{kk';q}(\mathcal{E}),
\end{eqnarray}
with the short-hand notations for function $\Pi_{kk';q}\left(\mathcal{E})=\Pi(\mathcal{E}+\varepsilon_q-\varepsilon_k-\varepsilon_{k'}-\frac{1}{2}\varepsilon_I(q-k-k')\right)$. The obtained equations clearly demonstrate that the trial wave function (\ref{dimer}) reproduces the diagrammatic calculations of the dimer energy in the $T$-matrix approximation \cite{Hryhorchak_2026}. Furthermore, the dimer is always energetically preferable over the trimer, whose energy, calculated in the $T$-matrix approximation, is given by the solutions to equation $\mathcal{T}^{-1}_{0;0}(\mathcal{E})=0$. The latter equation, however, does not properly account for particle-hole excitations, so the energies of the dimer and trimer states are calculated using completely different approximations. The same type of inconsistency in the $T$-matrix approximation regarding the description of the polaron-molecule transition is intrinsic to the Fermi-polaron problem, especially in lower dimensions \cite{Parish_2011}.

\subsection{Trimer state}
When the number $N$ of host particles is macroscopic, the simplest idea of a trimer suggests the bound state of two impurities and exactly one fermion outside the Fermi surface, composed now of the remaining $N-1$ particles. The trial wave function, which correctly embodies this idea and takes into account one particle-hole pair, should be chosen as follows:
\begin{eqnarray}\label{trimer}
&&|T\rangle=\sum_{k}t_ke^{-ikY}\psi^{\dagger}_k|N-1\rangle\nonumber\\
&&+\frac{1}{2L}\sum_{k,k',q}t_{kk';q}e^{i(q-k-k')Y}\psi^{\dagger}_k\psi^{\dagger}_{k'}\psi_q|N-1\rangle,
\end{eqnarray}
where a zero value of trimer momentum is assumed. A generalization of the above ansatz to nonzero center-of-mass momenta is straightforward. Conventional $T$-matrix results can be reproduced by taking into account only the first term in the trial wave function (\ref{trimer}). The evaluation of the expectation value $\langle T|H_{\textrm{eff}}|T\rangle$ is simple, although cumbersome. Then, by calculating two variational derivatives, one obtains a system of coupled integral equations (see Appendix B). After introducing additional notations $f_{k;q}\propto\frac{1}{L}\sum_{k'}t_{kk';q}$, the system reduces to one linear inhomogeneous integral equation
\begin{eqnarray}\label{trimer_Eqs1}
&&\mathcal{T}^{-1}_{k;q}(\mathcal{E}+\varepsilon_F)f_{k;q}+\Pi_{k;0}(\mathcal{E}+\varepsilon_F)\left[1+\frac{1}{L}\sum_{q'}f_{k;q'}\right]\nonumber\\
&&=\frac{1}{L}\sum_{k'}\Pi_{kk';q}(\mathcal{E}+\varepsilon_F)f_{k';q},
\end{eqnarray}
and one transcendental equation determining the trimer energy that should be satisfied simultaneously
\begin{eqnarray}\label{trimer_Eqs2}
\mathcal{T}^{-1}_{0;0}(\mathcal{E}+\varepsilon_F)=-\frac{1}{L^2}\sum_{k,q}\Pi_{k;0}(\mathcal{E}+\varepsilon_F)f_{k;q}.
\end{eqnarray}
Here $\varepsilon_F$ is the chemical potential of host fermions and we have utilized the short-hand notations for function $\Pi_{k;q}\left(\mathcal{E})=\Pi(\mathcal{E}+\varepsilon_q-\varepsilon_k-\frac{1}{2}\varepsilon_I(q-k)\right)$. Note that only non-vanishing terms in the limit $\Lambda \to \infty$ are preserved in Eqs.~(\ref{trimer_Eqs1}), (\ref{trimer_Eqs2}), and the trimer energy is again measured from the ground-state energy of $N$ non-interacting host fermions. These two equations fully determine the trimer state variationally in the one-particle-hole approximation. Comparing energies $\mathcal{E}$ calculated from Eq.~(\ref{Tau_3}) and from Eqs.~(\ref{trimer_Eqs1}),(\ref{trimer_Eqs2}) for arbitrary dimensionless couplings $\ln(|\epsilon_3|/\varepsilon_F)$ and mass ratios $m/m_I$, we can identify the phase diagram of the system.

\subsection{Results}
\begin{figure}[h!]
\includegraphics[width=0.235\textwidth]{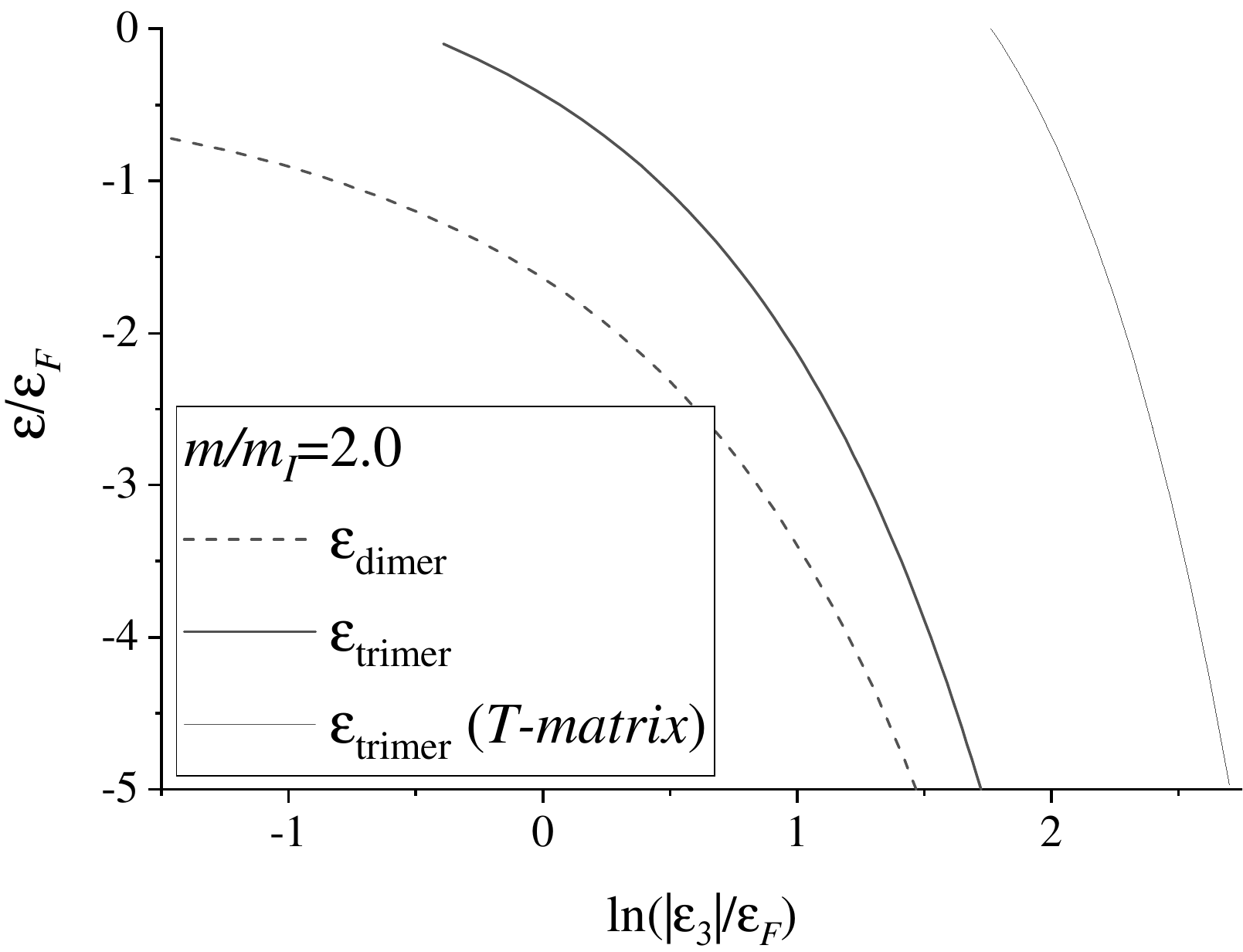}
\includegraphics[width=0.235\textwidth]{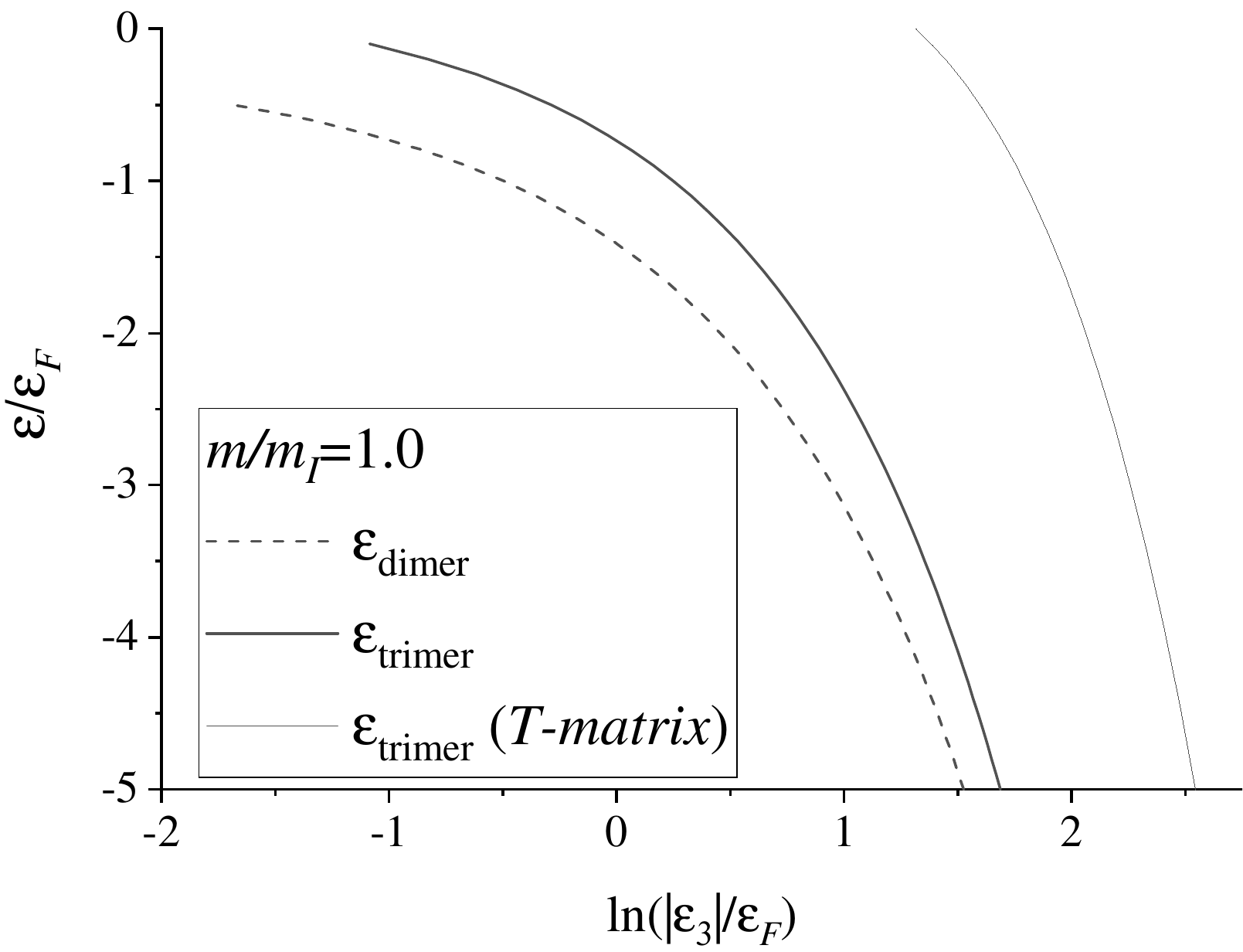}
\includegraphics[width=0.235\textwidth]{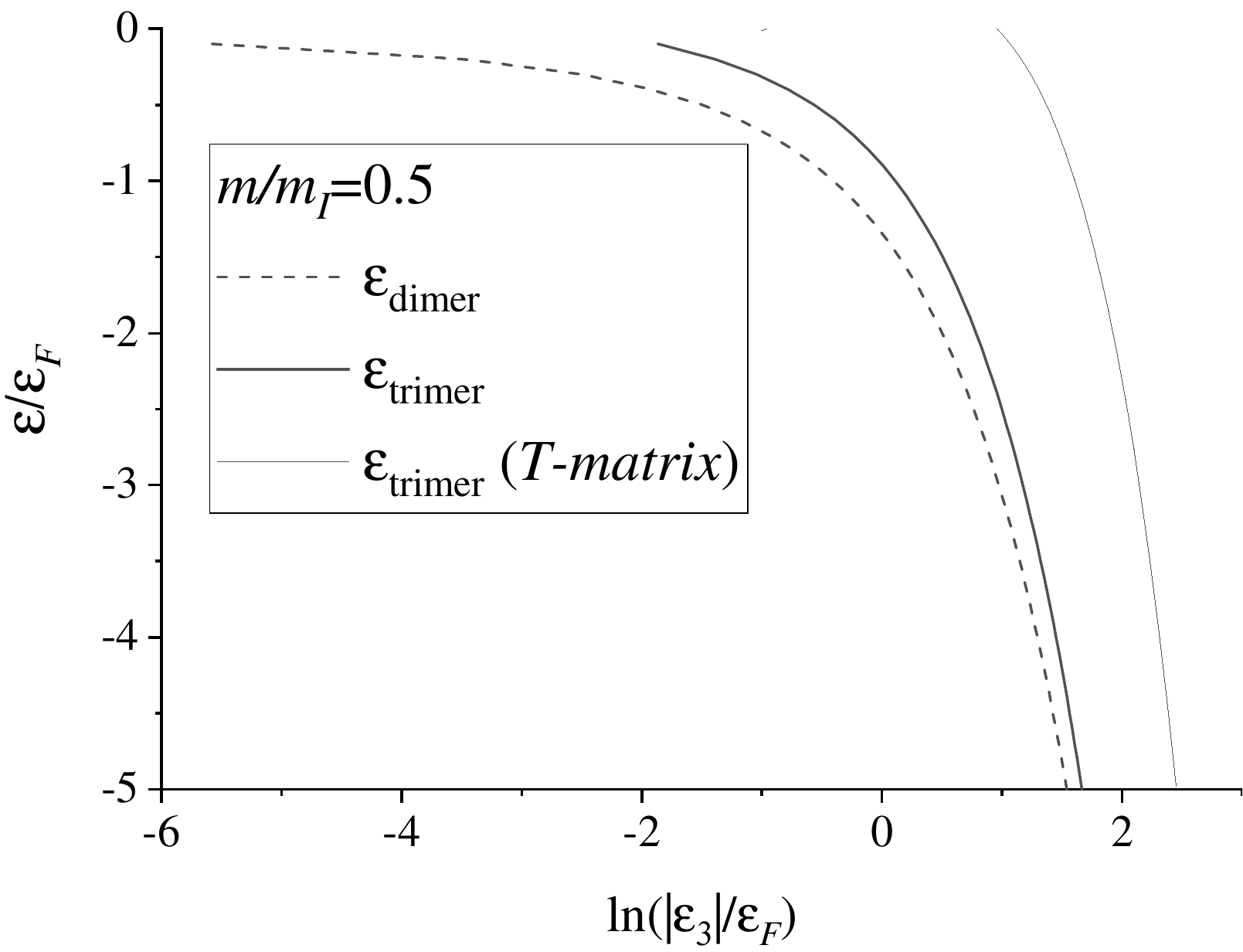}
\includegraphics[width=0.235\textwidth]{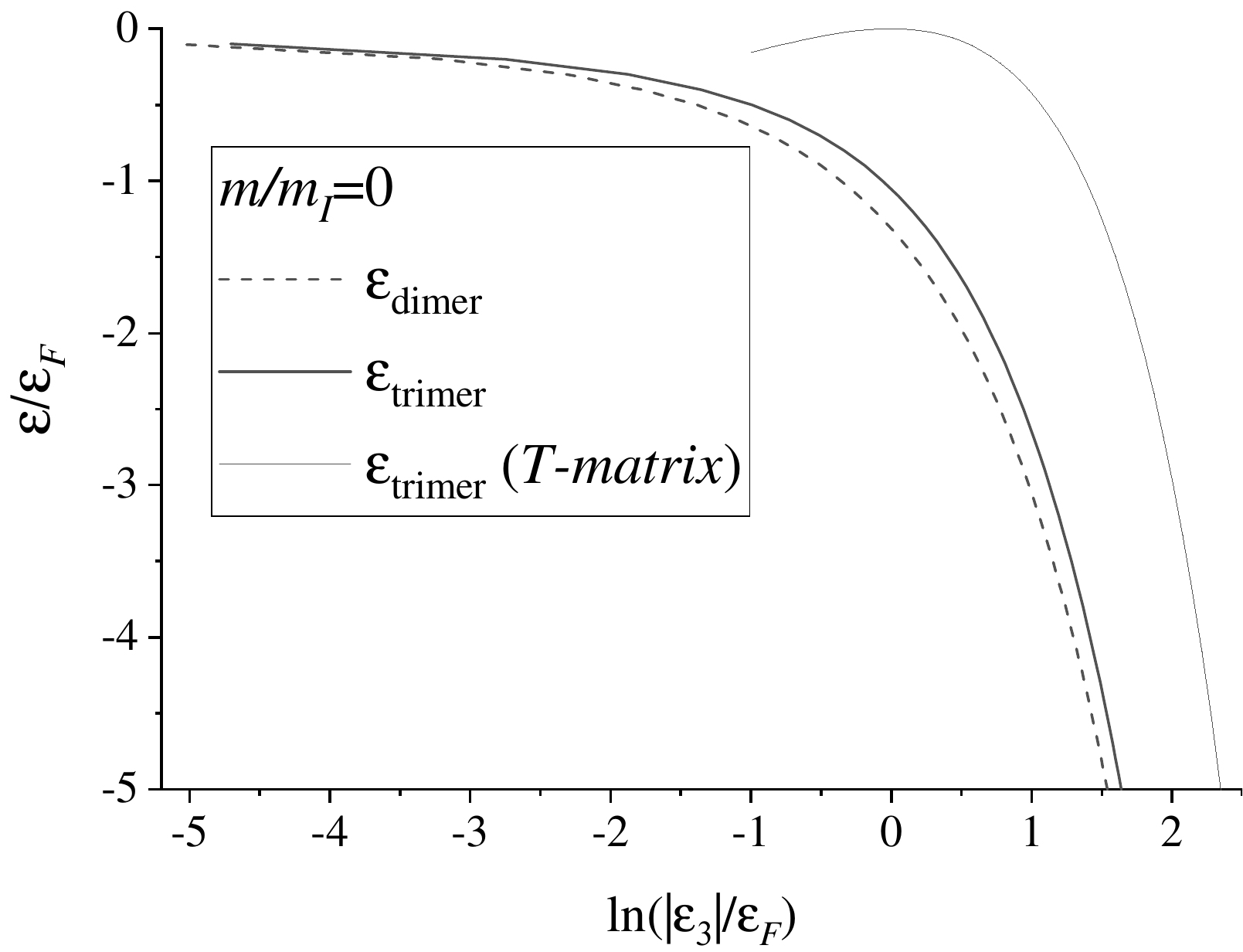}
	\caption{Energies of trimer (solid line) and dimer (dashed line) in the one-particle-hole approximation for different mass ratios $m/m_I$. The thing line represents the result of the $T$-matrix approximation for the trimer.}\label{trimer-dimer_fig}
\end{figure}

For the dimer state, we calculated the integral determining $\mathcal{T}_{0;q}$ explicitly via elementary functions and then performed numerical integration in the transcendental equation (\ref{dimer_Eq}). The dimer energies for different sets of parameters were obtained by numerically solving Eq.~(\ref{dimer_Eq}) using standard numerical Python methods. Calculations of the trimer properties are more cumbersome and require more computational time. First, in the thermodynamic limit ($L\to \infty$, $N\to \infty$ with $N/L$ kept fixed) we transformed the integral equation (\ref{trimer_Eqs1}), determining the function $f_{k;q}$ of two variables into a matrix equation [from a practical point of view, it is more natural to use a variable $1/k \in (-1/p_F,1/p_F)$ instead of $k$ and rewrite all integrations appropriately]. Then, at fixed $\ln(|\epsilon_3|/\varepsilon_F)$ and $m/m_I$, we solved the obtained inhomogeneous matrix equation for different energies $\mathcal{E}$, checking the fulfillment of the condition (\ref{trimer_Eqs2}). In Fig.~\ref{trimer-dimer_fig}
we present the results of these calculations alongside the corresponding dimer energies. The error bars are smaller than the symbol sizes.
To elucidate the role of the one particle-hole excitations, the results for the trimer energy are contrasted with the non-self-consistent $T$-matrix calculations
\begin{eqnarray}\label{E_T_matrix}
\frac{\mathcal{E}}{\varepsilon_F}=\frac{m}{2m_I}-\left(\sqrt{1+\frac{m}{2m_I}}-\sqrt{\frac{|\epsilon_3|}{\varepsilon_F}}\right)^2,
\end{eqnarray}
valid for negative $\mathcal{E}$s. In general, we observe a very unusual situation when the energy of the medium-induced two-impurity bound state (dimer) is energetically more advantageous over the vacuum-like quasiparticle (trimer) almost in the whole region of parameter space. Only in the limit of extreme diluteness $\ln(|\epsilon_3|/\varepsilon_F)\gg 1$ of host fermions, the trimer becomes more energetically preferable. Figures~\ref{trimer-dimer_fig} demonstrate a general tendency: the discrepancy of the dimer and trimer binding energies is less for massive impurities. On the other hand, it means that dimer-trimer transition happens for smaller $\ln(|\epsilon_3|/\varepsilon_F)$ if $m_I/m\gg 1$. Note that, in contrast to a standard Fermi polaron problem, here the limit of infinite impurity mass is not exactly solvable. 

\begin{figure}[h!]
\includegraphics[width=0.235\textwidth]{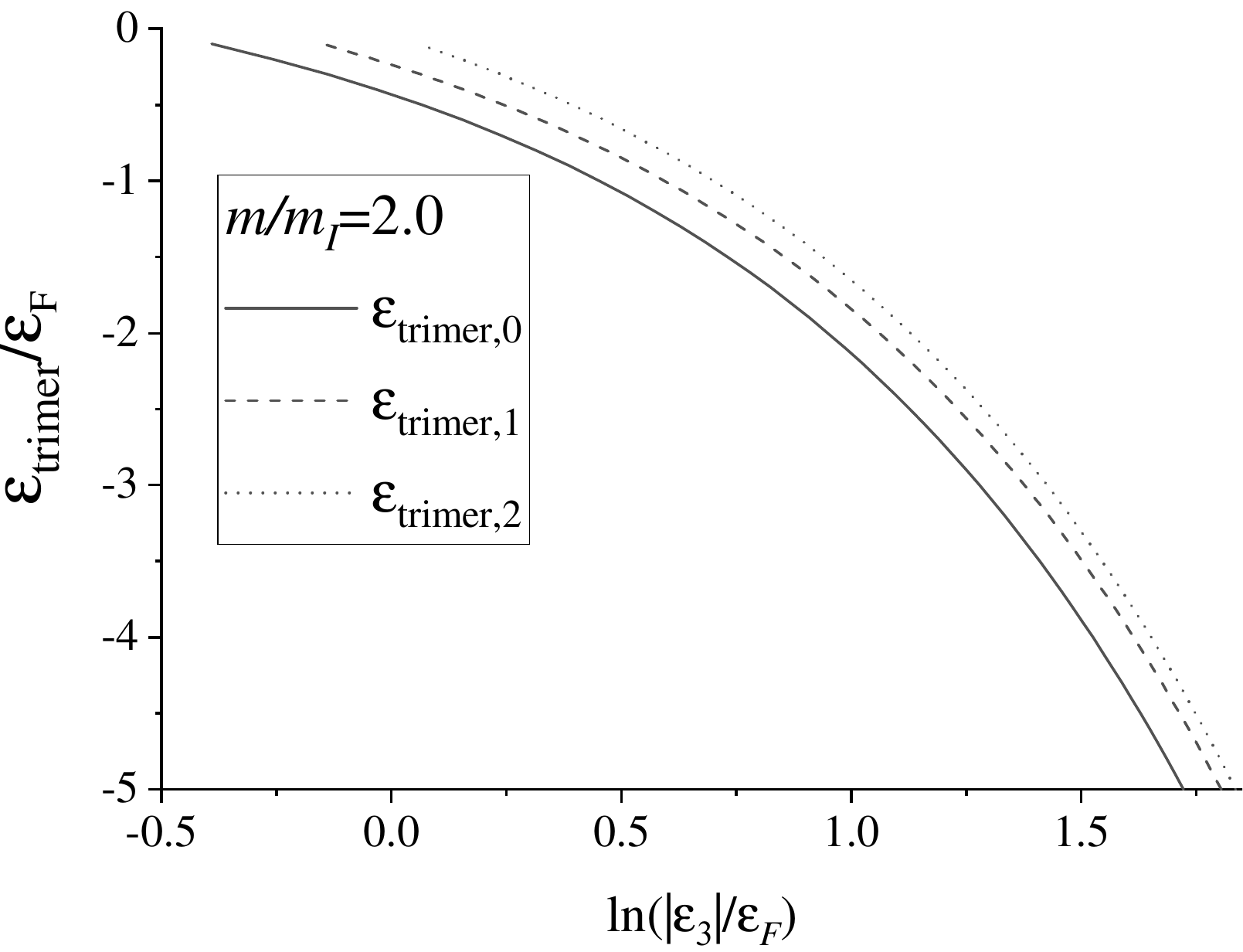}
\includegraphics[width=0.235\textwidth]{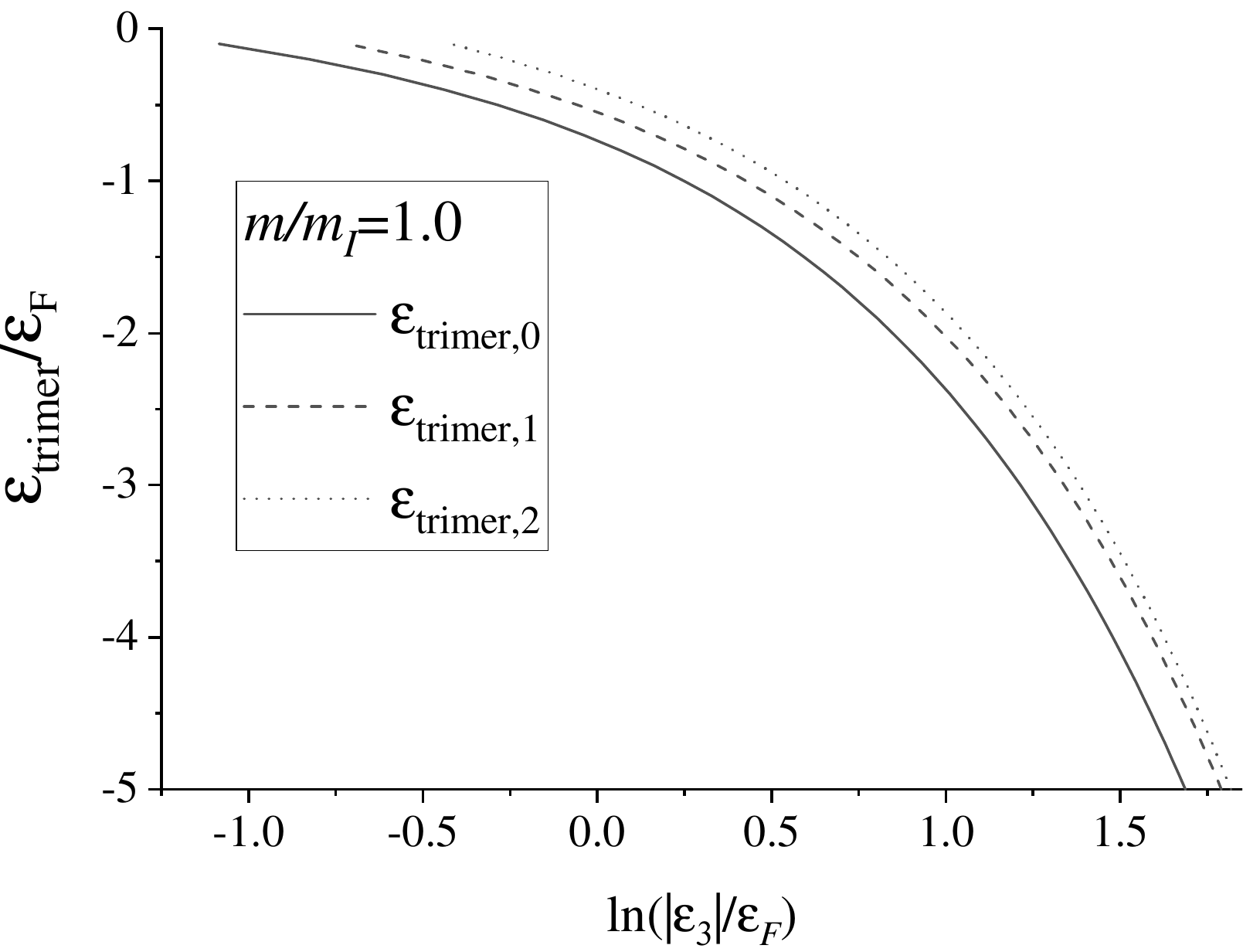}
\includegraphics[width=0.235\textwidth]{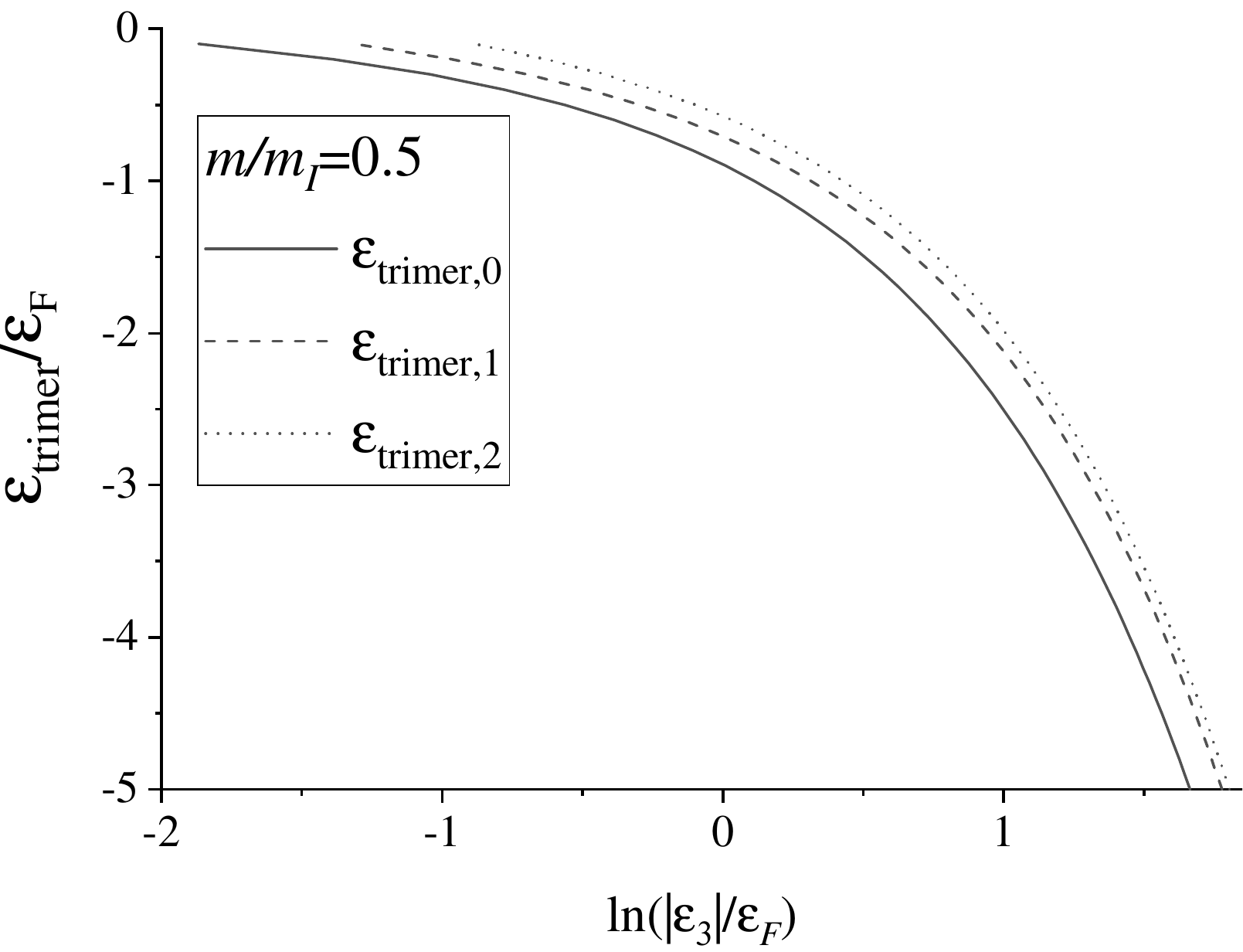}
\includegraphics[width=0.235\textwidth]{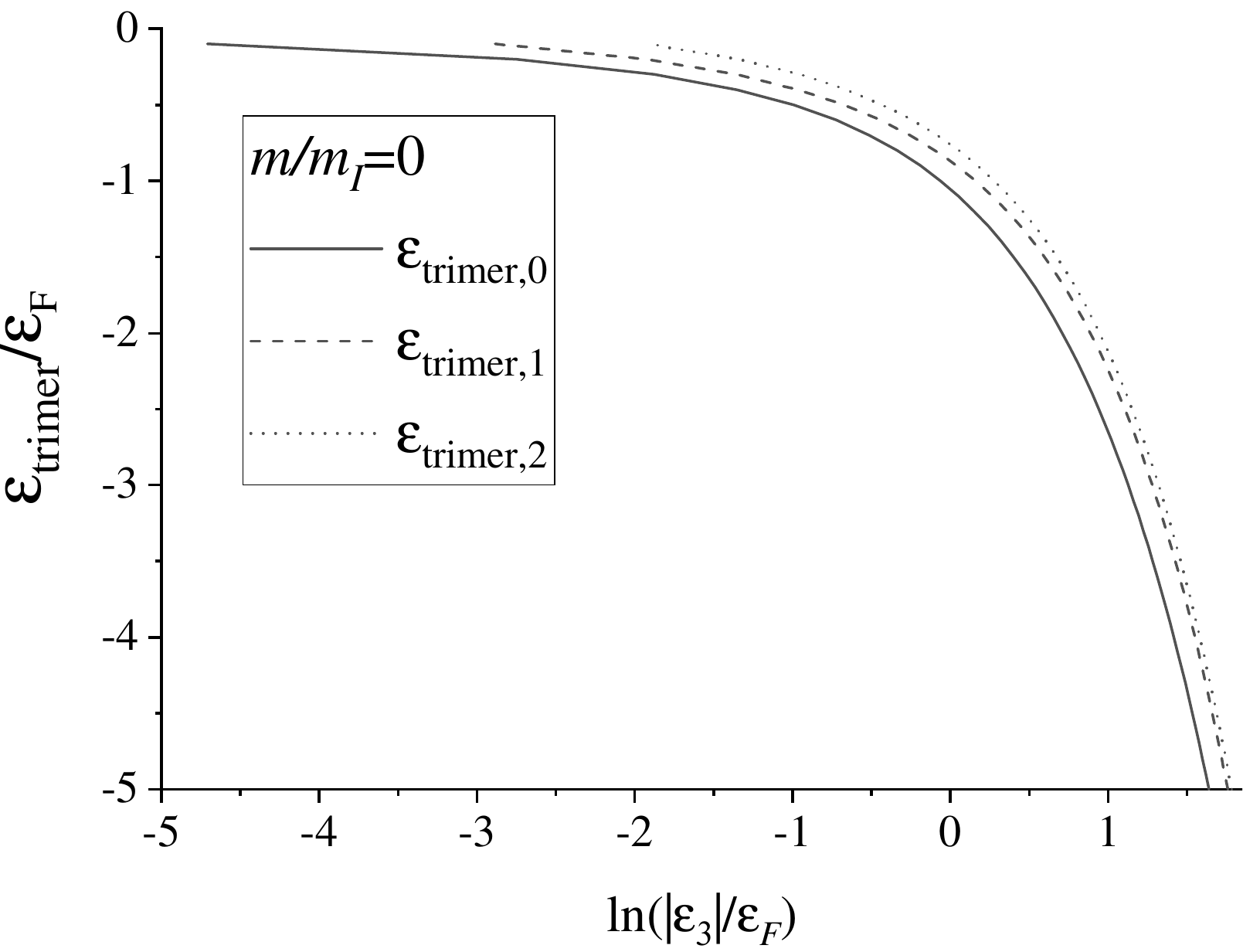}
	\caption{Identified branches of trimers for several mass ratios of impurities and medium fermions. The vacuum-like $\epsilon_{\textrm{trimer,0}}$ (solid line) is the lowest one that survives even in the three-body limit. Two others disappear at sufficiently low densities of host fermions.}\label{trimers_fig}
\end{figure}

Another important result of our calculations is that we found not a single trimer but rather a number of two-impurity-fermion bound state branches. Within an accessible computing power, we were able to identify two additional trimers reliably; however, there are some numerical hints for increasing this number. Their binding energy dependence on the coupling parameter for several mass ratios $m/m_I$ is presented in Fig.~\ref{trimers_fig}.
For comparison, we plotted the vacuum-like branch (solid line) announced in Fig.~\ref{trimer-dimer_fig}. It should be emphasized that the three-body contact interaction does not generate excited vacuum three-body states, and the emergence of extra trimers in the considered many-body system is a {\it fully collective effect}. Of course, at sufficiently low densities of medium fermions, all these excited trimers merge into the three-body continuum.

\section{Summary}
In conclusion, we have predicted rich medium-induced few-body physics in the one-dimensional Fermi bipolaronic system with a contact three-body interaction. It includes the trimer excited states and the two-impurity dimer, which are entirely due to collective effects. In particular, assuming that two impurities are immersed in a gas of spin-polarized fermions with all the two-body interactions suppressed, we have revealed the effect of the three-body interaction on the system's ground state. For these purposes, we have constructed the effective Hamiltonian that describes the interaction between the bipolaron center-of-mass coordinate and host fermions and exactly encompasses the properties of the system. Applying the variational principle that accounts for up to one particle-hole excitation, both for dimer and trimer states, we have constructed the phase diagram of the system. Independent of the magnitude of the fermion-impurity mass ratio, the medium-induced effective two-body attraction between impurities leads to the bound state formation, and this dimer prevails almost throughout the whole coupling parameter space. The dimer-trimer transition is realized only at extremely low densities of host fermions close to the three-body limit.

Unlike the standard Fermi-polaron problem, where the impurity interacts through a two-body potential, here all impurity-medium correlations originate from a contact three-body interaction. Consequently, the medium-induced attraction between impurities is generated through a qualitatively different microscopic many-body mechanism. The present variational theory can naturally be generalized to multiple particle-hole excitations, finite temperatures, and higher dimensions, opening a route toward the study of many-body phases of three-body Fermi bipolarons.

\begin{center}
	{\bf Acknowledgments}
\end{center}
We express our gratitude to participants of the 6th Conference ``Statistical Physics: Theory and Computer Simulations'' (August 27-29, 2025, Lviv, Ukraine) for discussing the results. This work was partly supported by Project No.~0122U001514 from the Ministry of Education and Science of Ukraine. The work of O.H. was supported by Project 2025.07/0326 (No.~0126U002943) from the National Research Foundation of Ukraine.

\section{Appendix}
This section discusses some details of the derivation of equations for dimer and trimer states.
\subsection{Dimer equation (\ref{dimer_Eq})}
The minimization of $\langle D|H_{\textrm{eff}}|D\rangle$ (at $P=0$) with respect to $d$ and $d_{k;q}$ leads to two coupled equations 
\begin{eqnarray}\label{}
\Pi^{-1}(\mathcal{E})d+g_{3,\Lambda}n+\frac{g_{3,\Lambda}}{L^2}\sum_{k,q}d_{k;q}=0,\\
\Pi^{-1}_{k;q}(\mathcal{E})d_{k;q}+g_{3,\Lambda}[d+nd_{k;q}]+\frac{g_{3,\Lambda}}{L}\sum_{k'}d_{k';q}\nonumber\\
-\frac{g_{3,\Lambda}}{L}\sum_{q'}d_{k;q'}=0,
\end{eqnarray}
where $n=\frac{1}{L}\sum_{q}$ is the density of host fermions. Introducing reduced amplitude $\bar{d}_{k;q}=d_{k;q}/d$ and using the second integral equation, we can construct the sum 
\begin{eqnarray}\label{}
\frac{g_{3,\Lambda}}{L}\sum_{k}\bar{d}_{k;q}=-\frac{\frac{g_{3,\Lambda}}{L}
	\sum_{k}\Pi_{k;q}(\mathcal{E})}{g^{-1}_{3,\Lambda}+\frac{1}{L}
	\sum_{k}\Pi_{k;q}(\mathcal{E})}
\end{eqnarray}
that remains finite after the limit $\Lambda\to \infty$ is carried out (all other terms equal zero in this limit and therefore have been dropped out).
Plugging it into the first equation of the system, we obtain Eq.~(\ref{dimer_Eq}).
  
\subsection{Trimer equations (\ref{trimer_Eqs1}),(\ref{trimer_Eqs2})}
The calculation and subsequent minimization of $\langle T|H_{\textrm{eff}}|T\rangle$  is a rather simple task, and we only write down linear equations determining amplitudes of the trimer state
\begin{align}
\Pi^{-1}_{k;0}(\mathcal{E}+\varepsilon_F)t_k+\frac{g_{3,\Lambda}}{L}\sum_{k'}t_{k'}+\frac{g_{3,\Lambda}}{L^2}\sum_{k',q'}t_{kk';q'}=0,\\
\Pi^{-1}_{kk';q}(\mathcal{E}+\varepsilon_F)t_{kk';q}+g_{3,\Lambda}[t_k-t_{k'}]+g_{3,\Lambda}nt_{kk';q}\nonumber\\
+\frac{g_{3,\Lambda}}{L}\sum_{k''}[t_{kk'';q}-t_{k'k'';q}]
-\frac{g_{3,\Lambda}}{L}\sum_{q'}t_{kk';q'}=0,\label{t_kkq}
\end{align}
note that $t_{kk';q}$ is an anti-symmetric function of $k$ and $k'$ by construction. Multiplying the first equation by $\Pi_{k;0}(\mathcal{E}+\varepsilon_F)$, performing the summation over $k$ and making use of notation $f_{k;q}=\frac{g_{3,\Lambda}}{L}\sum_{k'}\bar{t}_{kk';q}$ [where the rescaled amplitude $\bar{t}_{kk';q}=t_{kk';q}/\left(\frac{g_{3,\Lambda}}{L}\sum_{k''}t_{k''}\right)$; $\frac{g_{3,\Lambda}}{L}\sum_{k''}t_{k''}$ is finite at UV and determined by the wavefunction normalization], we can cast it into Eq.~(\ref{trimer_Eqs2}) in the main text. The normalized amplitude $\bar{t}_k=t_k/\left(\frac{g_{3,\Lambda}}{L}\sum_{k'}t_{k'}\right)$, then reads
\begin{eqnarray}\label{}
\bar{t}_k=-\Pi_{k;0}(\mathcal{E}+\varepsilon_F)\left[1+\frac{1}{L}\sum_{q'}f_{k;q'}\right].
\end{eqnarray}
In the same way, one can obtain Eq.~(\ref{trimer_Eqs1}) by expressing $t_{kk';q}$ from the second equation of the system with the subsequent summation over $k'$. Only the first, second, and fourth terms in Eq.~(\ref{t_kkq}) survive in the $\Lambda\to \infty$ limit, fixing a simple form
\begin{eqnarray}\label{}
\bar{t}_{kk';q}=-\Pi_{kk';q}(\mathcal{E}+\varepsilon_F)[f_{k;q}-f_{k';q}],
\end{eqnarray}
of the rescaled amplitude in the one-particle-hole approximation.

\end{document}